\documentclass[manuscript]{aastex}

\slugcomment{Submitted to PASP April 20, 2002}

\shorttitle{Rayleigh Laser Guide Star Systems:  Application to UnISIS}
\shortauthors{Thompson and Teare}

\begin{document}

\title{Rayleigh Laser Guide Star Systems:  Application to UnISIS}
\author{Laird A. Thompson\altaffilmark{1} and Scott W. Teare\altaffilmark{2,3}}
\altaffiltext{1}{Astronomy Department, University of Illinois Urbana-Champaign, 1002 W. Green St.,
Urbana, IL  61801; thompson@astro.uiuc.edu}

\altaffiltext{2}{Departments of Electrical Engineering and Physics, New Mexico Tech, 801 Leroy Place, 
Socorro, NM  87801; teare@ee.nmt.edu} 
\altaffiltext{3}{Astronomy Department, San Diego State University, San Diego, CA 92182}

\begin{abstract}Laser guide stars created by Rayleigh scattering provide a reasonable means to 
monitor atmospheric wavefront distortions for real-time correction by adaptive optics 
systems.  Because of the $\lambda^{-4}$ wavelength dependence of Rayleigh scattering, short 
wavelength lasers are a logical first choice for astronomical laser guide star systems, and 
in this paper we describe the results from a sustained experimental effort to integrate into 
an adaptive optics system a 351 nm Rayleigh laser guide star created at an altitude of 20 
km (above MSL) at the Mt. Wilson 2.5-m telescope. In addition to providing obvious 
scientific benefits, the 351 nm laser guide star projected by UnISIS is "Stealth qualified" 
in terms of the FAA and airplane avoidance.   Due to the excellent return signal at the wavefront 
sensor, there is no doubt that future applications will be found for short-wavelength Rayleigh scattered laser 
guide stars.
\end{abstract}


\section{INTRODUCTION}

UnISIS (University of Illinois Seeing Improvement System) is a laser guided adaptive 
optics system operating at the Coude focus of the 2.5-m telescope at Mt. Wilson 
Observatory.  It is the first astronomical system to employ a Rayleigh laser guide star 
at 351 nm.  While several descriptions of UnISIS have been published during system 
development and construction \citep{tho94,tho95,tho98}, this paper provides a complete over-view of 
the laser guide star system with detailed information on its design and its "as-built" configuration.  
The lessons learned in the UnISIS development effort will be of interest to those who are planning 
current generation laser guide star systems.  To keep this paper to a reasonable length but still 
describe key experimental issues in depth, only the UnISIS laser guide star system is 
discussed here.   Subsequent papers will describe the UnISIS adaptive optics system and 
its closed-loop performance characteristics including the cone effect (i.e., focal anisoplanatism) caused 
by the laser guide star's location in the near field when compared to astronomical objects.

Soon after the publication of the seminal paper describing the laser guide star concept by 
\citet{foy85}, \citet{tho87} reported on experimental work 
with laser guide stars and began detailed engineering design studies for laser guided 
adaptive optics systems (\citet{gar90} and references therein).  A 
key outcome of this early work was the realization that sodium laser guide stars, while 
conceptually attractive, would remain difficult to implement in the near-term especially if 
the goal is to successfully operate an adaptive optics system for scientific observations at 
wavelengths less than 2.2 microns.  Sodium wavelength lasers with sufficient power to 
adequately excite the sodium resonance line at 589.3 nm were not available at that time, 
and fifteen years later, these lasers are still difficult to obtain. To date, the only operational sodium
laser guide star system is that at Lick Observatory \citep{max97} where a 15 W sodium laser is used on 
a regular basis. The Lick Observatory laser -- and its ``sister'' laser at the Keck Observatory --
are one-of-a-kind systems built by Lawrence Livermore National Laboratory that will probably never be
duplicated again.

Rayleigh laser guide star techniques at 351 nm were discussed from the earliest times 
(\citet{tho89}; \citet{san94}) because commercial-quality excimer lasers capable of 
producing significant levels of pulsed power at 351 nm were already in production at that 
time.  This led to the experimental work of \citet{tho92} who reported on 
the creation and initial calibration of the return flux from a 351 nm laser guide star to 
altitudes up to $\sim$33 km.  After the Thompson and Castle experimental work had begun, the 
U.S. Air Force declassified information on the laser guided adaptive optics system at 
Starfire Optical Range \citep{fug91}.  The Starfire group had independently chosen 
to develop a Rayleigh laser guide star system that used the backscattered light from a 
copper-vapor laser (at 511 nm and 578 nm).  The Starfire system design \citep{fug94} -- even 
though independently devised -- closely matched the published design work 
by \citet{gar90}.  Subsequent visits to Starfire Optical Range 
provided information that improved the conceptual design of UnISIS. Every laser guided adaptive optics
system employs a unique set of tools to multiplex the optical system, to project the laser beacon
into the sky, to reject scattered light, etc. Our purpose here is not to present a comprehensive review
of these methods. Those interested in examples of other systems can refer to \citet{gre92}, \citet{fug94}, and 
\citet{san94}, and references therein.

Determining the ideal laser for Rayleigh scattered laser guide stars is the subject of 
another paper \citep{tho02a}, but it is worth noting here that short 
wavelength Rayleigh laser guide star systems are a viable alternative for astronomical 
purposes and that market forces continue to drive technological improvements for short 
wavelength commercial lasers.  UnISIS relies in a 30-Watt excimer laser originally built 
by Questek Incorporated.  This laser is no longer in production, its basic structure having 
morphed into the laser systems now used by VISX for LASIK eye surgery.  About two 
years ago Lambda Physik placed in production an excimer laser system more powerful 
($\sim$150 Watts) and better suited for astronomical use called Lambda Steel (developed for 
the manufacture of flat panel displays).  It is also worth noting that diode pumped and 
frequency tripled Nd:YAG and YLF lasers, which operate at 355nm and 349 nm, 
respectively, are also available commercially, and both are viable systems for short 
wavelength Rayleigh laser guide star systems.

This paper contains a diverse collection of both design information and practical 
experience obtained during the development of UnISIS.  It will provide assistance to those 
who are beginning the process of designing and implementing a laser guide star system.  
The field of laser guided adaptive optics is still in an early phase of development, and 
there are many new ideas to discover and to exploit.  In this paper, the most notable 
achievements are the "Stealth" characteristic of the UnISIS laser guide star system (Sec. 
5) and the successful acquisition of the laser wavefront return signal from the Rayleigh 
laser guide star (Sec. 8).

\section{UnISIS CONFIGURATION AT MT. WILSON OBSERVATORY}

The UnISIS deformable mirror and adaptive optics system sit on a fixed optics bench at 
the Coude focus of the Mt. Wilson 2.5-m telescope, and the excimer laser is housed in a 
special air conditioned room on the observatory ground floor 9.5 m below the Coude 
room.  Figure 1 shows a schematic drawing of this arrangement.  The Coude system at the 
2.5-m telescope is the simplest possible consisting of only three mirrors:  the primary, 
secondary, and tertiary, with the tertiary located at the intersection of the right ascension 
and declination axes of the telescope mount.  Because the UnISIS adaptive optics system 
sits at Coude, there is essentially no flexure nor misalignment among the adaptive optics 
components themselves as the telescope tracks an object across the sky.  This design also 
provides ample space for an open and flexible optical design in contrast to the alternate 
design of placing the adaptive optics system at the Cassegrain focus.

The input beam from an equatorially mounted telescope rotates at the sidereal rate when 
the beam is sent onto a fixed optics table at the Coude focus, so the telescope pupil image 
-- with the shadow of the secondary mirror support struts -- rotates on the fixed UnISIS 
deformable mirror.  Indeed, the telescope beam rotates on all fixed elements along the 
UnISIS beam line, and the final science focal plane would also rotate if countermeasures 
were not taken.  While it is possible to remove this beam rotation at the interface between 
the telescope and the Coude optics bench (with a reflective "dove prism" rotator), this 
option was not adopted, and corrective action is taken only at the final science focal plane.  
The visual wavelength science CCD is on a mechanical rotation stage, and images 
collected at the near-IR science detector are image post-processed to remove field 
rotation.

\subsection{Subsystem Drift and Alignment Criteria}

UnISIS consists of three subsystems:  the laser itself with its projection optics, the 
adaptive optics bench, and the 2.5-m telescope system.  These three subsystems have to be 
co-aligned to high precision.  Four degrees of freedom are required to co-align one optical 
system relative to another; these represent a fixed x, y position of the optical axis and the 
two direction cosines of the beam.  The tip-tilt controls on a pair of mirrors will suffice to 
link one subsystem to another.  Figure 2 shows a side-view of the two-mirror beam 
transfer system that carries the telescope Coude optical axis onto the UnISIS optics bench.  
As the figure shows, these two mirrors are relatively close to the Coude focus.  A similar 
two-mirror interface links the laser projector to the system optical axis.

The three main subsystems of UnISIS drift in time with respect to one another with 
amplitudes that are large enough to affect system performance.  When UnISIS is prepared 
for operation, the main optics table is taken as the primary reference, and the other two 
sub-systems are brought into co-alignment with this reference.  At the image scale of the 
f/30 Coude focus (2.75 arcsec/mm), x, y position alignment to 0.25 arcsec requires a 
precision of 90 microns, a specification that is not difficult to achieve.  The angular co-
alignment specification between subsystems is set by the strict acceptance angle criterion 
of the high-speed shutter that sits in front of the laser guide star wavefront camera; this 
vector co-alignment must be better than one part in $10^5$ $\sim$2 arcsec.  During initial 
UnISIS tests, and before the alignment procedure became standard, the return laser guide 
star signal would appear and disappear depending on whether this angular alignment was 
within the $\sim$2 arcsec range or not.  This is no longer an issue.

Current alignment procedures are routine and take approximately 20 to 30 minutes at the 
beginning of an observing session.  Realignment is generally required on a daily basis, but 
if day-to-day temperature changes inside the telescope dome are small, the system 
alignment will hold.  As we best understand, the subsystem drifts are caused by the 
following effects.  First, the ground floor laser room is on a solid foundation and is 
unlikely to shift.  However, the UnISIS Coude optics table rests on a metal framework 
bolted to the cement shell that forms the enclosure for the original Coude spectrograph.  
This 10-m high cement shell slowly rises, falls and twists with changes in the temperature.   
No doubt, a similar differential motion of the telescope optical axis occurs when the 
telescope's two steel piers expand and contract with ambient temperature changes.  These 
changes occur on $\sim$12 hour timescales and have little impact on the performance of 
UnISIS during a single night.

\section{RAYLEIGH LASER GUIDE STAR SYSTEM DESIGN}

\subsection{Basic Considerations}

The return flux from a Rayleigh laser guide star depends on atmospheric molecular and 
aerosol backscatter coefficients as well as on the absorption coefficient for the wavelength 
of interest.  The integrated atmospheric absorption (from the ground to the altitude of the 
laser guide star and which includes low altitude Rayleigh scatter) enters the relationship as 
a square because it acts in both the up-link and 
down-link laser paths.  These radiative transfer details will be described elsewhere 
\citep{tho02a}, and here it is sufficient to say that photons at 351 nm are 
ideal for creating laser guide stars because the Rayleigh scattering coefficient is high due 
to the $\lambda^{-4}$ wavelength dependence, and the absorption coefficient is relatively low at 351 
nm.  Ozone becomes a strong absorber only at wavelengths shorter than 345 nm.

For Rayleigh beacon systems, the laser must be pulsed to allow range gating, which also 
provides a "heart-beat" to coordinate the system timing.  Immediately following the laser 
pulse, strong Rayleigh backscattered light from low and intermediate altitudes must be 
rejected by a high-speed electronic shutter system capable of opening during the precise 
interval when the backscattered return signal arrives from the high altitude. Continuous 
lasers can be used for Rayleigh laser guide stars but only if they are mechanically 
chopped.

The optical system used to transmit laser light to altitude can be one of two types:  (1) a 
small collimator that transmits a uniformly narrow column up through the atmosphere, or 
(2) a large focusing element that transmits a converging beam focused to a specific 
altitude (see \citet{foy85}, \citet{gar90}, \citet{har98}, 
and references therein). While this choice is normally made based on the laser beam quality 
-- a poor laser beam quality requiring a large focusing element -- to make a laser transmitter 
``Stealth qualified'' a high quality laser beam could be transmitted with a large focusing element.
Note that UnISIS employs the second of these two options with the 2.5-m primary mirror 
acting as the focusing element because the laser beam quality is poor.  The UnISIS laser beacon is 
focused $\sim$18 km above the 
telescope.  Because Mt. Wilson Observatory is located 1.8 km above sea level, the laser 
guide star return signal comes from an atmospheric layer $\sim$20 km above mean sea level.  
A schematic representation of the laser beacon projection is shown in Figure 3.

\subsection{Integrated Laser Guide Star Depth}

The light from the outgoing laser beam fills the conical volume shown in Figure 3.  The 
base of the cone is at the primary mirror of the 2.5-m telescope and its tip sits at a distance 
of 18 km.  Once the laser light reaches the tip of the cone, it passes through a beam waist 
and fills an inverted conical volume that starts at 18 km.  The usable laser guide star return 
signal comes from the double conical volume limited by the lines labeled $\Delta$$z_t$ and $\Delta$$z_b$ in 
Figure 3.  The total length $\Delta$z for optimal return is derived in \citet{tho89} to be

\begin{displaymath}
	\Delta z  =  ( \Delta z_t + \Delta z_b ) = 4.88 \lambda z_o^2  / ( D_p  r_o ),
\end{displaymath}

where all symbols in this equation are defined in Fig. 3 except $\lambda$ which is the wavelength of 
the laser transmission and $r_o$, the Fried seeing cell size for the laser projection wavelength.  This 
criterion for $\Delta$z assumes that the limit for the lateral dimension of the time-gated laser 
return signal (as viewed from a distance $z_o$) equals the atmospheric seeing size for the 
laser wavelength.  Note the inverse dependence of $\Delta$z on $D_p$ so that a smaller laser beam 
footprint on the telescope primary mirror allows a deeper illuminated volume to contribute 
to a seeing-limited laser guide star image.  The Thompson and Gardner relationship was 
devised for a telescope with no central obscuration, so a point of diminishing returns will 
occur with UnISIS if (to gather more return flux) $D_p$ is decreased to the point where it 
begins to approach the diameter of the Cassegrain secondary mirror.  For UnISIS 

\begin{displaymath}
		 \lambda = 351 nm,\;        z_o = 18 km,\;      D_p \sim 1.8 m,\;    r_o = 14.2 cm,
\end{displaymath}

where the value for $r_o$ is scaled to the laser wavelength but is otherwise taken from the 
500 nm median value of $r_o$ as reported by Walters and Bradford (1997) for Mt. Wilson 
and implies that the median Mt. Wilson seeing is between 0.7 and 0.8 arcsec at visual 
wavelengths.  In these conditions, the predicted depth for the integrated laser volume is $\Delta$z  
= 2.2 km.  The value used above for $D_p$ is discussed below. Additional information on the 
astronomical seeing at Mount Wilson is provided in \citet{tea00a,tea02} that show that there 
are long-term trends in the seeing profile. 
		
\subsection{Questek Excimer Laser}

Figure 4 shows the Questek 2580v$\beta$ excimer laser system used to produce the UnISIS 
laser guide star.  This laser operates in a pulsed mode with pulse length $\sim$20 ns, repetition 
rates programmable up to 500 Hz, and an average power of approximately 30 Watts.  
[Each laser pulse is $\sim$90 mJ, and since the laser power scales directly with the repetition 
rate, 30 Watts is the nominal laser power for the 333Hz UnISIS mode of operation.]  The 
output beam quality is poor compared to other laser systems, but beam quality can be 
corrected with appropriate optical components as described below.  The Questek 2580v$\beta$ 
was purchased in 1990 and was used in both the \citet{tho92} Rayleigh 
guide star experiments and the laser guide star work of \citet{ney02} before being 
moved to Mt. Wilson.  

When plane-parallel laser mirrors are mounted at both ends of the laser chamber, the 
natural divergence of the excimer laser beam is $\sim$3 milliradians, approximately $10^3$ 
larger than required for creating a laser guide star close to the seeing limit.  Better 
performance is obtained in two steps.  First, the planar laser mirrors are replaced with a 
pair of curved windows which form a miniature Cassegrain telescope.  The rear laser 
mirror acts like a primary and the front laser mirror acts like a Cassegrain secondary 
mirror.  In this configuration the laser is said to have an unstable resonator cavity.  Fig. 5 
shows the cavity and the photon amplification paths.  This laser amplification proceeds as 
follows.  Seed photons in the core of the laser beam that happen to be traveling in the 
forward direction hit the convex front laser mirror and travel in the backwards direction 
through the laser chamber.  These backwards traveling photons are amplified and expand 
to fill the rear laser mirror.  The rear laser mirror is concave and is designed to collimate 
photons reflected off the front laser mirror.  In the final forward pass through the lasing 
chamber, these amplified seed photons sweep a large fraction of the energy from the 
lasing medium and emerge from the front laser window as the output pulse.  In the process 
of beam expansion within the laser chamber by the Cassegrain optical system, the laser 
beam divergence is reduced from 3 milliradians to approximately 300 microradians 
\citep{cau99}.  If this beam were transmitted into the upper atmosphere without further 
modification, the laser would produce a small focused spot approximately 62 arcsec 
across.  This beam divergence is still too large by a factor of 100 to create a useful laser 
guide star.

\subsection{Full-Aperture Laser Projection}

The output beam from the Questek laser is approximately 9 mm x 22 mm, and it is this 
beam that has a nominal divergence of 300 microradians.  If a small divergent beam of 
diameter d is expanded to fill a collimator of diameter D, the divergence in the final beam 
is reduced by the factor D / d.  For UnISIS the focusing element is the telescope's primary 
and secondary mirror.  If we adopt d = 24mm for the circle that encloses the raw laser 
beam and D = 2.5 m for the largest focusing element, the divergence is 2.9 microradians 
(0.6 arcsec) after it is beam-expanded and projected into the sky off the 2.5-m primary 
mirror.  

This logic sets the design strategy for projecting the UnISIS laser beacon into the sky.  
The true situation is somewhat more complex for the following reasons:  (1) the laser 
beam is masked and only part of the 9 mm x 22 mm makes it to the telescope pupil, (2) 
the remaining laser beam is expanded to fill only a central 1.8 m portion of the primary 
mirror (see below) and (3) the unstable resonator optics in the Questek 2580v$\beta$ do not 
produce a perfectly collimated output beam.  Instead, the laser beam has an empirically 
measured beam divergence of 620 microradians which consists of the quadrature sum of 
the random divergence (300 microradians) and a "mechanical" or "optical" divergence 
(540 microradians).  The imperfect optical divergence of the laser is easily corrected 
with the laser projection optics (described in section 4.3 below).

Experimental tests of laser projection from Mt. Wilson confirm that laser guide stars as 
small as $\sim$0.8 arcsec FWHM can be produced under good seeing conditions as long as the 
integrated depth of the laser guide star $\Delta$z is kept below 2.2 km as calculated above. 
Although Mount Wilson Observatory is not longer a dark site \citep{tea00b}, the U band 
sky brightness does not degrade the laser return signal because the laser guide star sits 
significantly above the sky background light equivalent to a 9th or 10th magnitude star.

\section{BEAM-SHARING AND SYSTEM TIMING}

After the 351 nm beam exits the laser on the observatory ground floor and passes into the 
Coude room, it is converted into an f/30 to f/40 diverging beam so that it can join the 
telescope optical axis and be projected into the sky.  The major design complication is 
how to allow the laser projection system to beam-share the optical axis with the incoming 
astronomy light.  In the laser guided adaptive optics system at Starfire Optical Range, the 
beam-sharing element was a solid cube beam splitter \citep{fug94} that had the 
unfortunate property of fluorescing because of the very high flux of laser light that passed 
through it.  This fluorescence produced a relatively continuous flux of background light 
that prevented the detection of low surface brightness astronomical objects.  

\subsection{Rotating Glass Disk}

To avoid fluorescence, the UnISIS beam-sharing scheme consists of a rotating disk with 
small reflective spots deposited on its front surface.  The laser light is directed towards the 
rotating glass disk (at an angle of incidence of $\sim$22 degrees), and at the exact moment 
when the reflective spot is sitting on the telescope optical axis, the laser is fired.  Once the 
laser fires,  the outgoing laser pulse hits the small reflective spot and is sent into the sky 
along the telescope's optical axis.  As quickly as the light travels up to $\sim$18 km and back, 
the small reflective spot moves off the optical axis and the laser guide star light traces a 
reversed path, passes through a clear area on the rotating disk, and proceeds into the 
UnISIS adaptive optics system.  

Astronomy light from the sky passes through the rotating disk at all times and into the 
adaptive optics system.  The fraction of lost astronomy light (reflected from the spots) is 
at most $\sim$5.5\% (the area obscured by the spots).   Because these spots are multi-layer 
dielectric coatings, they appear nearly transparent at visible wavelengths, so the amount of 
light lost to the science cameras is likely to be less than 2\%.   Figure 6 shows a picture of the 
rotating disk assembly sitting on a table after being removed from the optical system.  The 
disk itself is 0.25-inch thick UV-grade fused silica with excellent surface flatness and an 
outer diameter of 5 inches.  Multi-layer dielectric reflective spots (R = 99.9\% at 351 nm) 
are deposited in sets of three (the three-spot configuration is explained in Thompson and 
Xiong 1995) at radii of 41.6, 45.6, and 49.6 mm.  These spots are hardened to withstand 
laser powers as high as 4 to 6 J $cm^{-2}$.

Glass disks of diameter 5 inches have sufficient strength to withstand the rotational stress 
of 10,000 RPM, and the disk shown in Figure 6 rotates at this rate.  The rectangular 
reflective spots are 3.4 mm in the radial direction and either 4.2 mm (for the inner spot) or 
4.7 mm (for the two outer spots) in the azimuth direction.  The reflective spot size is set 
by the requirement that they must rotate out of position fast enough to allow the return 
light from altitude not to be obscured when it returns from $\sim$18 km altitude.  For the 
UnISIS disk specifications, this requirement is met for Rayleigh backscatter altitudes 14.5 
km and higher, a range that corresponds to the low edge of the laser guide star range gate 
($\Delta$$z_b$ in Fig. 3).  

If just one reflective spot were deposited on the disk, at 10,000 RPM it would appear on 
the optical axis 167 times per second, and this would set the frequency or "heart-beat" of 
the wavefront sensing in UnISIS.  Because closed-loop performance of an adaptive optics 
system might suffer if it were run at this rate, pairs if spots are deposited on the disk at 
180 degree separations which allows 333 Hz adaptive optics operation.  For redundancy 
sake, four sets of spots are deposited on the disk (at 90 degree spacing) as a fall-back 
should any of the spots be damaged by the high laser energy density (which indeed has 
happened).

Because the laser light reflects off the front surface of the rotating disk, a tight 
specification had to be set for the extent to which the motor shaft is perpendicular to the 
front surface of the disk.  Otherwise, in 333 Hz operation, the angle of incidence for the 
reflected beam would be different for every other spot, and the laser guide star at altitude 
would move from side to side every other pulse.  The mechanical specification for the 
disk relative to its shaft was set at less than 3 arcsec.  The acquisition of the UV grade fused silica 
disk was relatively simple, but getting the glass disk attached to the motor shaft to the 
required level of precision was more difficult.

\subsection{UnISIS System Timing}

UnISIS requires the synchronization of three subsystems to complete one cycle of operation. First, 
the spots on the rotating disk must be synchronized with the arrival of the outgoing laser energy 
so that the 351nm light is directed out the telescope. Second, the Pockel's cell must be opened at
the appropriate time to receive the return signal. Third, the wavefront sensing camera must be triggered
to receive the return pulse from the 18km focus.

The master clock for this system is the rotating disc controller, a stand-alone embedded microprocessor system,
which reads a motor encoder on the rotating disc and maintains its rotation speed at 10kHz. This same encoder
information also provides an indicator giving the position of reflective spots. When the spots are in the correct 
orientation, it sends a TTL pulse to the laser system. In order to trigger the laser precisely (to within 
10 ns accuracy) the laser is left in its fully charged configuration, waiting for a pulse from the rotating 
disc controller.  The rotating disc controller has external dipswitches used to manually enter a phase delay 
between the rotating disc spot positions and the laser trigger pulse.  The rotating disk is viewed in a 
stroboscopic mode in order to set the phase delay and get the laser to fire when a reflective spot is exactly 
on the optical axis.  

Immediately after the laser fires, an optical fiber photovoltaic converter (placed at the edge of the outgoing laser 
beam) senses the outgoing laser pulse and generates a TTL pulse.  This TTL pulse is sent to a 
digital delay / pulse generator (Stanford Research Systems Model DG535) which is 
connected by BNC cables to the high voltage drivers for the Pockel's cell switch and to an 
external "strobe" line into the camera electronics of the laser guide star wavefront sensor.  
By dialing appropriate time delays into the digital delay generator, the Pockel's cell 
shutter is opened and closed at the appropriate times (i.e. the speed of light travel time to $z_o$ 
- $\Delta$$z_b$ and $z_o$ + $\Delta$$z_t$) and for triggering the exposure of the wavefront sensor CCD.  
The CCD camera then passes a data frame to the wavefront reconstructor, a separate standalone computer with 8 
internal digital signal processors connected in parallel. The reconstructor calculates the wavefront corrections 
to be passed to the 177 actuator Xinetics deformable mirror electronics and a single cycle of wavefront correction ends.
UnISIS is currently run at three repetition rates:  17 Hz for testing and system set-up, 167 
Hz operation mode to match nights when atmospheric wavefront timescales are slow, and 
the 333 Hz mode to match nights when atmospheric timescales are fast.

\subsection{f/30 Laser Projection Optics}

As mentioned above, the laser reflective spot must rotate off the optical axis rapidly 
enough to allow the return Rayleigh light to pass through a clear area on the rotating disk.  
This requirement, and the fact that glass disks will not withstand arbitrarily high rotation 
rates, means that the reflective spot must be small and that the laser beam must be reduced 
to a small dimension before it encounters the disk.  This adds just a few complications to 
the system design:  the beam from the laser must be reshaped anyway from its very gentle 
540 microradian divergence to match the $\sim$f/30 Coude beam.  This reshaping is done in 
the Coude room in the vicinity of the 18 km conjugate point, a point that is located 328 
mm beyond the Coude infinity focus.  The details of this beam reshaping were first 
discussed by \citet{tho95}.

In quick summary, the laser light passes along the following path.  The beam emerges 
from the front laser port in the basement laser room, hits three flat beam-directing mirrors 
(used for sub-system co-alignment) and travels vertically for $\sim$11 m where it encounters 
an adjustable razor blade pupil stop.  At this point the laser beam has expanded to 14 mm 
x 28 mm, but the razors are used to reduce the beam to 14 mm x 24 mm to avoid spillover 
onto the rotating disk beyond the reflective spot.  The clipped beam then passes through 
an air-spaced doublet that forms a relatively fast converging beam as shown in Figure 7. Just before this beam 
reaches focus, it enters a small diverging lens that converts the clipped laser beam into the 
expanding $\sim$f/30 beam that hits the rotating disk and travels up the telescope south polar 
axis.  The beam then hits the tertiary, secondary, and primary mirrors and is transmitted to 
the 18 km focus.  Figure 8 shows two images of the laser beam, one in its full rectangular 
form and the other clipped and on its way to up the telescope's south polar axis.

The three-element UV-grade fused silica optical system is designed to create a virtual 
image of the laser guide star 70 mm behind the back surface of the small diverging lens.  
By placing these three optical elements in a position such that the virtual image of the 
laser guide star sits at the conjugate focus of the 18 km layer (namely 328 mm beyond the 
infinity focus), the combined optical system that includes the telescope primary mirror 
creates a concentrated region of laser energy some 18 km above the primary mirror with a 
waist $\sim$52 mm in diameter (2.9 milliradians as seen from the primary mirror) with 
additional smearing caused by atmospheric blurring.  The Zemax optical design for this 
laser projection system (including the 2.5-m telescope optics) is given in Table 1.

The air-spaced doublet in the 3 element optical system is mounted in a cell that can be 
moved along the optical axis while the diverging lens remains fixed.  As the doublet 
moves, the output f/ratio and the location of the virtual image conjugate point change 
simultaneously.  By watching the laser return signal with a test camera focused on the 18 
km layer (see Sec. 7 below), the position of the air-spaced doublet can be optimized for 
the best focus at 18 km.  In the most recent experimental tests, the optimal laser guide star 
focus occurred at a somewhat slower output beam than expected ($\sim$f/39) and hence the 
laser pupil does not completely fill the primary mirror.  This has the effect of slightly 
increasing the focal spot size at 18 km from the ideal 52 mm value given above, but this 
does not degrade the performance significantly because the limiting angular size of the 
laser guide star is dominated by atmospheric blurring in the roundtrip up-link and down-link 
paths.  The fact that the beam is being transmitted at $\sim$f/39 means that the laser beam 
footprint on the primary mirror is somewhat less than the full 2.5-m aperture and hence 
the value given above for $D_p$ $\sim$1.8 m.

The alignment and positioning of the three-element laser projection optical system is very 
critical to good performance of UnISIS.  The better focused the laser guide star is, the 
more accurately the Shack-Hartmann wavefront sensor can determine the wavefront error.  
The process of focusing the laser guide star is somewhat dangerous because the energy 
density in the converging beam -- at the point just before it enters the small diverging lens 
-- becomes quite large.  Slight misadjustments in the positions of the three projection 
lenses can cause the energy density to rise above the damage thresholds of the optical 
coatings.  Coatings on these lenses (and on the multiplayer dielectric spots on the rotating 
disk) were applied by Acton Research Corporation (Acton, MA) and have damage 
thresholds of 4 - 6 J cm$^{-2}$.  One accident did occur where the converging beam was 
allowed to shrink too much, and it exceeded damage threshold given above.  Figure 9
shows the results.  Newer diverging lenses have now been installed in UnISIS with 
dielectric coatings applied by Alpine Research Optics (Boulder, CO).  These have a 
damage threshold if 17.5 J cm$^{-2}$.

\section{FAA AND LASER SAFETY ISSUES}

When compared to other laser guide star systems, UnISIS has a clear advantage in being 
generally benign to both aircraft and satellite interference.  In this regard, UnISIS might 
be called "Stealth qualified".  Because this characteristic affords great convenience in the 
nighttime operation of at Mt. Wilson, it is appropriate to describe how this was achieved.  
The situation begins with the fact that 351 nm light is invisible to the eye, and it is further 
aided by the need to for full-aperture laser projection as explained in sections 3.3 and 3.4 
above.  Full-aperture laser projection significantly dilutes the laser beam intensity at all 
altitudes except in the focused beam waist at $\sim$18 km. Laser guide star projection systems 
that project at visible wavelengths and those that use a relatively small collimators are 
required by the FAA to implement safety procedures that include airplane spotters located 
outside the telescope dome.

The surface energy density of the UnISIS laser beam as it leaves the telescope primary 
mirror is very dilute.  A person who might accidentally stand in the beam at a point 
immediately outside the telescope dome would receive a continuous UV flux 
approximately equal to the UV flux one would receive from the Sun in the same 350 nm 
region of the spectrum.  351 nm light -- from the Sun or from a laser -- is of sufficiently 
low energy that it does not damage proteins in the body so it will not lead to the formation 
of cancer, but it can penetrate the aqueous humor of the eye and cause cataracts, so laser-safe 
glasses are used by the UnISIS team and the telescope operators who are in the dome 
on nights when the laser is being transmitted.  Laser-safe eyeglasses or goggles are 
required in the laser room and in the Coude room where the laser beam is hot and the 
scattered UV photon flux is substantial.  

When the telescope is pointed to the zenith, the UnISIS laser beam comes to a waist at 
20 km above mean sea level, which corresponds to 66,000 ft.  Because seeing degrades 
as a function of zenith distance and because differential refraction (between the UV laser 
guide star and the near-IR wavelengths of the science cameras) increases as a function of 
zenith angle, there are no plans to use UnISIS further than 40 degrees from the zenith.  
Therefore, the minimum altitude for the laser guide star beam waist is 15.3 km (50,0000 
ft.), so commercial aircraft are not likely to enter the focused region of the laser guide star.  
However, at the altitudes of air traffic the energy flux (in J cm$^{-2}$ per pulse) in the 
laser beam will be greater than that immediately in front of the primary mirror, but to 
counter this point we note that photons at 351 nm are absorbed by the glass and the 
plastics that are used in aircraft windows.  

In a white paper sent to the FAA in 1997, the UnISIS laser was called a "single pulse 
laser" from the perspective of an airplane pilot or passenger.  The maximum operating 
pulse rate of the UnISIS laser is 333 Hz.  If we assume an airplane window is 30 cm 
across, an airplane would have to be flying slower than 30 m/s (67 mph) for two pulses to 
enter the window.  This is much slower than the speed of an average airplane.  ANSI 
standards list specifications for single pulse lasers in terms of the power per pulse at a 
given wavelength, and the dilute beam of the UnISIS laser -- in the altitude range where 
commercial planes fly -- places the UnISIS laser with Class I systems (eye safe), even if 
the laser photons were able to pass through glass or plastic windows.  One of the main 
worries of the FAA is the distractive nature of a laser beam in the sense that a pilot notices 
the emission.  Since 351 nm light is invisible to the eye, this is not an issue either.

After initial telephone contact with the Western-Pacific Region of the Federal Aviation 
Administration in Los Angeles, CA, and after explaining the circumstances described 
above, the FAA requested a written report restating the case.  Approximately 4 months 
after the FAA received the UnISIS report, they issued a letter which stated:  

"Propagation of Class I lasers into navigable airspace pose no hazard to aviation and we 
have no objection to the operation described in your correspondence.  Please advise our 
office of any proposed changes or alteration to your proposed scientific research laser 
installation in order that an aeronautical study may be accomplished to determine the 
effect of the changes on the safe and efficient use of navigable airspace by aircraft."

\section{OPTICAL ALIGNMENT AND FOCUS METHODS}

\subsection{Sub-System Co-Alignment}

As described above (sec. 2.1), there are relative drifts over $\sim$12 hour periods between the 
three UnISIS subsystems.  This persistent problem made it necessary to understand and 
carefully control the UnISIS optical alignment.  This exercise came to a successful end 
after a few hard rules were imposed.  These included (1) to define with great precision a 
single vector on the UnISIS adaptive optics bench as the primary system reference, (2) to 
align each subsystem separately and only then pull them together as a whole, and (3) 
master the chromatic problems associated with the alignment of refractive optics that 
work at 351 nm but must be aligned -- at least initially -- at the more convenient 635 nm 
laser diode wavelength.  These issued are discussed in turn.

The single reference vector on the UnISIS adaptive optics bench is defined with two 
mechanical reference points.  One of these is a small (but removable) pinhole that sits in 
the center of the primary telescope focal plane (at the f/30 infinity focus).  The second 
reference point is a removable point-like target located 1.5 meters beyond the infinity 
focus. A small but very bright diffraction-limited laser diode beam is forced through these 
two mechanical points no matter where the UnISIS optics bench happens to sit as it 
"floats" on its underlying support structure.  This reference beam is used to co-align the 
individual components in the adaptive optics section of UnISIS and then the same is done 
for all components in the laser guide star projection subsystem.  Except for chromatic 
effects discussed below, this procedure works well.  Then a second reverse-traveling 635 
nm diode laser reference beam is co-aligned with the first, and this second beam is 
propagated up through the Coude optical train of the telescope.  Using the tip-tilt action of 
the two mirrors shown schematically in Fig. 2, the telescope sub-system is finally brought 
into co-alignment with the other sub-systems.  As noted above, the procedure defined in 
this paragraph takes approximately 20 to 30 minutes to complete.

The chromatic alignment problems in the refractive sections of UnISIS were the most difficult to 
understand and defeat.  There are two refractive systems within UnISIS:  (1) the laser 
guide star projection system (described in Sec. 4.3 above) and (2) the UV wavefront 
sensor optics that include the Pockel's cell shutters and the EEV wavefront sensor camera 
(described in Thompson, Teare , Crawford and Leach 2002).  One very convenient 
procedure for aligning a lens-based system is to send a narrow pencil-beam laser through 
the optical train and search for faint retroreflected and focused ghost images that are 
formed by each curved (lens) surface.  These focused ghost images sit at 1/2 x the radius 
of curvature of each optical surface.  By identifying the positions of both ghost images 
(one on the front side and one on the back side of a standard double convex lens), the tip-tilt 
and the x,y centroid for a single lens can be set, one lens at a time.  While this is a 
straightforward procedure, even the slightest angular misalignment between the 635 nm 
reference beam and the true 351 nm entrance vector can produce significant errors in the 
alignment, especially if there are multiple lenses in the system because the position of the 
635 nm alignment vector gets displaced, one lens after the next, relative to the 351 nm 
vector by refractive effects within each lens.  The only way to remove these ambiguities is 
the use 351 nm light as the final check on the alignment.  While 351 nm photons are 
difficult and inconvenient to use in alignment exercises (because they are not visible to the 
human eye), 351 nm photons are indirectly detectable because they produce a blue 
fluorescence on a high-cotton-content paper target.  Careful cross checks of the 635 nm 
alignment with 351 nm laser light made it possible to eliminate the chromatic effects in 
the refractive sections of UnISIS.

\subsection{Laser Guide Star Focus and Alignment}

The laser guided adaptive optics system built at Starfire Optical Range \citep{fug94} introduced
the important concept of calibrating the laser guide star wavefront signal during closed-loop 
operation on a natural star. As described in another paper \citep{tho03}, UnISIS incorporates 
this design concept from Fugate et al. Although it requires simultaneous operation of two wavefront 
cameras -- one for a natural star and the other for the laser guide star -- this technique 
unambiguously removes the non-common path aberrations that inevitably exist between the laser guide
star wavefront optical path and the natural star optical path. In UnISIS the natural star wavefront
sensor -- used for this calibration -- is situated immediately before the final science camera focal
plane. By running UnISIS in closed-loop on a natural star and (in open-loop) recording a 
Shack-Hartmann wavefront reference frame for the laser guide star, this calibration technique delivers
to the wavefront computer system what we call the ``laser wavefront Shack-Hartmann reference frame''. 
Only after recording this calibration frame is UnISIS run in closed-loop with the laser guide star. The wavefront computer is 
instructed to drive the deformable mirror not to the mechanical null of the laser guide star Shack-Hartmann
sensor but to the null of the``laser wavefront Shack-Hartmann reference frame''. Even if there are uncalibrated optical
offsets in the laser guide star optical system (e.g. problems of focus or astigmatism), the natural 
star calibration method circumvents these problems and the adaptive optics system drives to a null representing
the diffraction limited performance of the natural star calibration. In this way we avoid the need 
to establish the absolute true focus and alignment of the laser guide star wavefront system.  

\section{TIME-GATED IMAGE TUBE CAMERA}

During the UnISIS system development, a time-gated image intensifier camera was used 
repeatedly to identify and diagnose problems with the laser guide star system alignment 
and focus.  There was simply no other convenient means available to detect the faint high 
altitude laser guide star return signal in the presence of the low altitude and very bright 
foreground Rayleigh scattered light.  When the CCD wavefront sensor was used without 
its Pockel's cell shutter, it was quickly driven into saturation by the high-intensity but 
low-altitude Rayleigh scattered light.

The time-gated Reticon camera functions essentially the same as the primary UV wavefront sensor
system, with the following modifications. Instead of sending the TTL output 
of the digital delay generator to the Pockel's cell switches, this signal is sent to the high 
voltage gated input line of the image intensifier.  When the intensifier is off, the Reticon 
sensor is blind (to the low altitude Rayleigh flash).  For laser guide star tests, the 
intensifier is turned on only for the short time interval (13 $\mu$s to 17 $\mu$s long) when the 
backscattered photons are returning from high altitude.  Even though the image intensifier 
is non-linear and must be handled cautiously in the presence of the bright outgoing laser 
pulse, a tool of this kind is invaluable to those who contemplate developing a Rayleigh 
laser guide star system.

\section{SHACK-HARTMANN WAVEFRONT TESTS}

The UnISIS UV wavefront sensor is designed to work in coordination with the UnISIS 
Rayleigh laser guide star system to produce Shack-Hartmann signals to drive the UnISIS 
deformable mirror.  Prior to the laser guide star tests described below, operation of the 
wavefront sensor was validated in the following two ways:

(1) An artificial laser guide star was sent through the UnISIS optical system by mounting 
a UV transmitting fiber optic on an x-y-z mechanical stage and positioning the fiber at the 
18 km conjugate point on the UnISIS optics bench.  The input end of this fiber was fed 
351 nm light from the Questek laser in such a way that the output end of the fiber optic 
produced a uniformly illuminated f/30 output beam.  With the dome closed and with the 
Questek laser placed in full operation, an external pulse generator was used to trigger the 
laser, the Pockel's cell shutter and the CCD camera to produce test wavefront images.

(2) On an excellent photometric night, the star Sirius was acquired at the normal infinity 
f/30 Coude focus of the telescope.  A focus screen was then installed at the 18 km 
conjugate focus (328 mm beyond the infinity focal plane), and the telescope secondary 
mirror was adjusted to bring the starlight into focus at the 18 km conjugate.  After 
removing the focus screen, the light from Sirius was passed through the adaptive optics 
system and onto wavefront sensor camera.  In this case, the CCD camera and the Pockel's 
cell shutters were run with an external pulse generator, and the CCD exposure time was 
set to 100 $\mu$s (the maximum exposure available with the camera drive electronics).  A 10 
nm wide interference filter centered at 350 nm was placed in the collimated beam 
immediately before the wavefront camera.  Test exposures from the wavefront sensor 
were then taken of Sirius.

The wavefront camera system worked as designed for both tests, and then UnISIS was 
prepared for the Rayleigh laser guide star tests.  Nights selected for these tests were of 
reasonable photometric quality so a return flux comparison could be made with Sirius..  
Only nights of excellent seeing were used for these tests because the individual pixels in 
the UnISIS UV wavefront sensor have dimensions of 1x1 arcsec.  The aim of the test was 
to focus the laser guide star light on single pixels on the wavefront sensor camera in 
order to estimate the FWHM of the laser guide star and to measure
the strength of the return laser guide star signal.  For these tests the 
telescope was parked within 1 or 2 degrees of the zenith.

Normally, the lenslet array in a Shack-Hartmann sensor is positioned for maximum 
wavefront sensitivity so that Hartmann subimages fall in the center of (at the central 
intersection of) a 2x2 pixel "quadcell".  For the test shown here, the 13x13 lenslet array 
was offset from this nominal position in two regards.  First, the lenslet array was moved 
significantly to the lower left (as seen in the two test images reproduced in Figure 10) to 
pass light to the sensor without going through the lenslet array.  Second, the lenslet array 
was positioned so that light from single lenses in the lenslet array were centered on a 
single pixel in the wavefront sensor.  This alignment was set immediately before the observing
run began, and based on previous experience, we know that the alignment will remain stable for 
many hours.

Wavefront data shown in Figure 10 were taken on the night of 5 Nov. 2001 (UT).  On this 
night the laser guide star system was operated in the standard test mode with a repetition 
rate of 17 Hz.  The UnISIS reconstructor computer has the ability to direct to a special 
storage buffer (on external command) 45 consecutive wavefront data frames.  For this test, 
the time gate on the Pockel's cell shutter was adjusted in real time to optimize the focus 
and the brightness of the lenslet images as viewed through a continuous video output from 
the UnISIS wavefront reconstructor computer.  Note that the laser guide star projection 
optics were placed at a particular mechanical point set by inspecting the return laser guide 
star image on the time gated Reticon camera; and no effort was made to iterate between 
the best appearing Pockel's cell setting and this mechanical focus.  The Pockel's cells are 
driven by a TTL pulse (from the Stanford delay generator) and have an 8$\mu$s rise-time after 
the trigger signal is applied to their high voltage power supplies.  The two time gates 
given below are for the time interval at half-maximum transmission and the time interval 
for 100\% Pockel's cell transmission:

    $t_b$ = 109 $\mu$s,   $t_t$ = 134 $\mu$s,   $\Delta$z = 3.75 km,   (50\% Pockel's cell transmission)

    $t_b$ = 113 $\mu$s,   $t_t$ = 130 $\mu$s,   $\Delta$z = 2.55 km,   (100\% Pockel's cell transmission)

Figure 10 (left) shows the wavefront image from a single pulse of the laser guide star.  Notice 
that several individual pixels are brightly illuminated whereas others are much fainter.  
This is the expected behavior for a Shack-Hartmann sensor running in open-loop, i.e.  
wavefront gradients from the turbulent atmosphere momentarily displace individual 
subimages from their nominal position, in this case centered on single pixels.  The three 
brightest pixels in Figure 10 (left) have 27$e^-$, 25$e^-$, and 25$e^-$ whereas the pixels immediately
adjacent to the three brightest pixels contain an average of 4$e^-$. This shows that the Rayleigh laser
guide star had a FWHM less than or equal to 1 arcsec (a limit set by the dimensions of a single pixel) 
on the night of 5 Nov. 2001.  Figure 10 (right) shows an integrated 
sum of 45 consecutive wavefront images.  The action of the turbulent atmosphere in this 
case is to blur the subimages, but their centroids, on average, remain centered on the 
individual CCD pixels.  The total detected signal from the individual bright pixels plus the 
flux in the 4 adjacent pixels (top, bottom, left, right) is 23$e^-$.

The UnISIS wavefront sensor used in these tests is a Marconi EEV-039a with a 
quantum efficiency of 48\% and a system read noise of 4.8 $e^-$ rms.  A new Marconi EEV-
039a sensor was recently obtained with 68.2\% quantum efficiency, and it is expected to 
have a comparable, or perhaps even lower, read noise.  The new sensor returns a signal 
that is higher by a factor of 1.42.  At the time the exposure in Figure 9 was taken, the 
excimer laser had been running several hours and the power per pulse was 75 mJ, whereas 
under normal conditions the laser runs at 90 mJ per pulse.  Therefore, the UnISIS 
wavefront camera under normal operating conditions with the new sensor will be 
detecting $\sim$40$e^-$ per subaperture at an rms read noise $\sim$4.8$e^-$ (or better).  This 
performance is perfectly satisfactory for closed loop laser guide star operation of UnISIS.

\section{CONCLUSION}

The UnISIS Rayleigh laser guide star system has been commissioned and operated in 
open loop with satisfactory return signal for a 13 x 13 set of subapertures across the pupil 
of the Mt. Wilson 2.5-m telescope.  A tightly focused laser guide star return signal with a 
FWHM $\sim$1 arcsec was received from a $\Delta$z = 2.5 km range gate centered at 18.2 km above 
the telescope ($\sim$20 km above mean sea level).  This successful demonstration of the 
UnISIS laser guide star system sets the stage for closed loop operation with the full 
UnISIS adaptive optics system.  The 351 nm Rayleigh laser guide star technique with full-
aperture broadcast provides a reasonable method to monitor wavefront perturbations 
in the Earth's atmosphere, and it is especially attractive because of its "Stealth" 
characteristics.  Rayleigh laser guide stars are likely to be the basis for many other 
laser guided adaptive optics applications in the future.

\acknowledgments
A number of people have contributed to the work reported here.  This includes Richard 
Castle, Dr. E. Harvey Richardson, Samuel Crawford, Dr. Xiong Yao-Heng, Dr. Chris 
Neyman, and Bill Knight and his machine shop crew.  At Mt. Wilson Observatory we 
acknowledge support provided by Mt. Wilson Institute Director Dr. Robert Jastrow and 
technical assistance by Robert Cadman, Sean Hoss, Chris Hodge, Joe Russell, Victor 
Castillo and Thomas Schneider of Schneider Engineering.  Telescope operator support at 
the 2.5-m telescope was provided by Kirk Palmer, Michael Bradford, and Jim Strogen.  
Information and assistance with the Questek laser system was provided by Dr. George 
Caudle and Geoffry Bramhall.  EEV wavefront camera electronics support was provided 
by Dr. Robert Leach, Jamie Erickson, and Scott Striet.  Assistance with the rotating disk 
assembly was provided by Geoff Gretton.  Advice was also provided by the UnISIS / NSF 
Advisory Panel that included Drs. Robert Fugate, Doug Simons, Matthew Mountain, 
David Sandler, Byron Welsh, Ray Weymann, G. Wayne van Citters, and Ben Snavely.  
This Rayleigh laser guide star research has been supported by grants from the National 
Science Foundation:  AST-8918878, AST-9220504, AST-0096741 and by funds from 
both the University of Illinois and the New Mexico Institute of Mining and Technology.  
All support is very gratefully acknowledged.

\clearpage

\begin{deluxetable}{crrrrr}
\tabletypesize{\scriptsize}
\tablecaption{Laser Guide Star Projection System:  Zemax Optical Specification. \label{tbl-1}}
\tablewidth{0pt}
\tablehead{
\colhead{Surface} & \colhead{Description} & \colhead{Curvature\tablenotemark{a} (1/r)} & \colhead{Thickness\tablenotemark{a}} 
& \colhead{Semidiameter\tablenotemark{a}} & \colhead{Glass}
}

\startdata

0  &\nodata            &0          &35400       &\nodata   &\nodata 
\\
1  &Laser Aperture     &0          &11400       &13.2      &\nodata  
\\
2  &1st Lens           &0.014422   &7.82        &17.45     &Silica  
\\
3  &\nodata            &0          &3.19        &16.99     &\nodata  
\\
4  &\nodata            &0          &\nodata     &16.59     &\nodata    
\\
5  &2nd Lens           &0.0186     &5.52        &16.28     &Silica    
\\
6  &\nodata            &0.0257     &\nodata     &15.16     &\nodata    
\\
7  &\nodata            &0          &150         &15.39     &\nodata 
\\
8  &Changed            &0          &30.68       &3.96      &\nodata
\\
9  &Diverging Lens     &0          &8.0         &1.63      &Silica   
\\
10 &Changed\tablenotemark{b}        &0.158      &$-$70       &1.21      &\nodata  
\\
11 &Virtual Focus      &0          &70          &0.037     &\nodata  
\\
12 &\nodata            &0          &50          &1.207     &\nodata   
\\
13 &Rotating Disc      &0          &$-$0.014    &2.04      &\nodata    
\\
14 &Changed            &0          &208         &2.04      &\nodata  
\\
15 &Inf Tel Focus      &0          &1100        &5.517     &\nodata   
\\
16 &Flat Mirror        &0          &0           &23.89     &Mirror    
\\
17 &\nodata            &0          &$-$8993.0   &23.89     &\nodata   
\\
18 &\nodata            &0          &0           &174.13    &\nodata  
\\
19 &Tertiary Mirror    &0          &7768.4      &174.13   &Mirror     
\\
20 &Secondary Mirror\tablenotemark{c}  &0.000137  &$-$7768.4    &304.02   &Mirror       
\\
21 &Tertiary Obscur    &0           &$-$2100    &330.2    &Obscuration   
\\
22 &Primary Mirror\tablenotemark{c} &3.879       &9868.4      &1270     &Mirror       
\\
23 &Loc of Secondary  &0           &17990000    &1293.6   &Obscuration  
\\
24 &Laser Star    &0           &0           &7.59     &\nodata 

\enddata 

\tablenotetext{a}{Units are mm.}
\tablenotetext{b}{70mm reversal to locate laser guidestar virtual image position.}
\tablenotetext{c}{Primary mirror conic constant is $-$1; secondary mirror conic constant is $-$1.979.}
\tablecomments{In this design the laser is modeled as a point source located
35.4 m behind the laser aperture, a distance selected to correspond to a beam divergence 
of 0.020 degrees. While this divergence was chosen somewhat arbitrarily, the value (and the
distance to the pseudo laser point source) do not significantly affect the design.}
\end{deluxetable}

\clearpage
\begin{figure*}
\epsscale{0.5}
\plotone{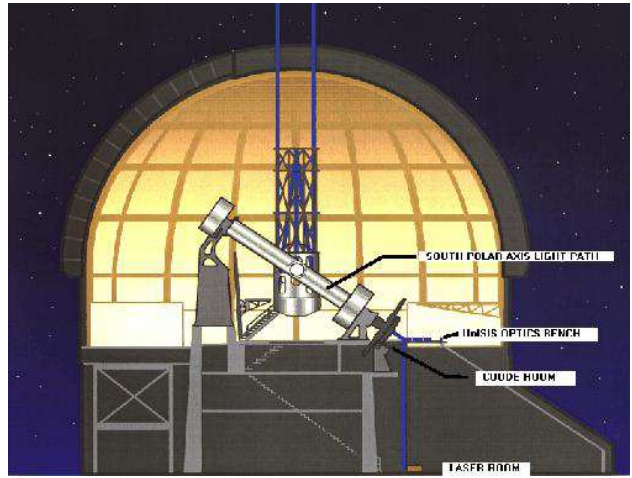}
\caption{Schematic drawing of the Mt. Wilson 2.5-m telescope and dome showing the 
location of the laser room, the UnISIS adaptive optics bench, the Coude room, and the 
Coude beam path along the south polar axis of the telescope.\label{fig1}}
\end{figure*}

\clearpage
\begin{figure*}
\epsscale{0.5}
\plotone{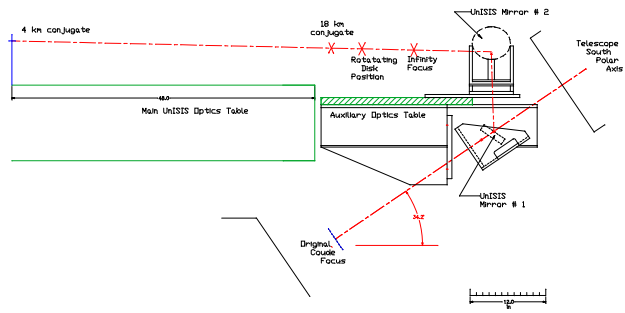}
\caption{Schematic drawing showing the Coude light 
path coming down the south polar axis of the telescope and being redirected by UnISIS 
Mirrors number 1 and number 2 onto the main UnISIS optics table.  The two mirrors are separated 
laterally (perpendicular to the plane of the drawing) by approximately 30 cm.   UnISIS 
Mirror number 2 is shown in a dashed outline because this view shows its back side.
\label{fig2}}
\end{figure*}

\clearpage
\begin{figure*}
\epsscale{0.5}
\plotone{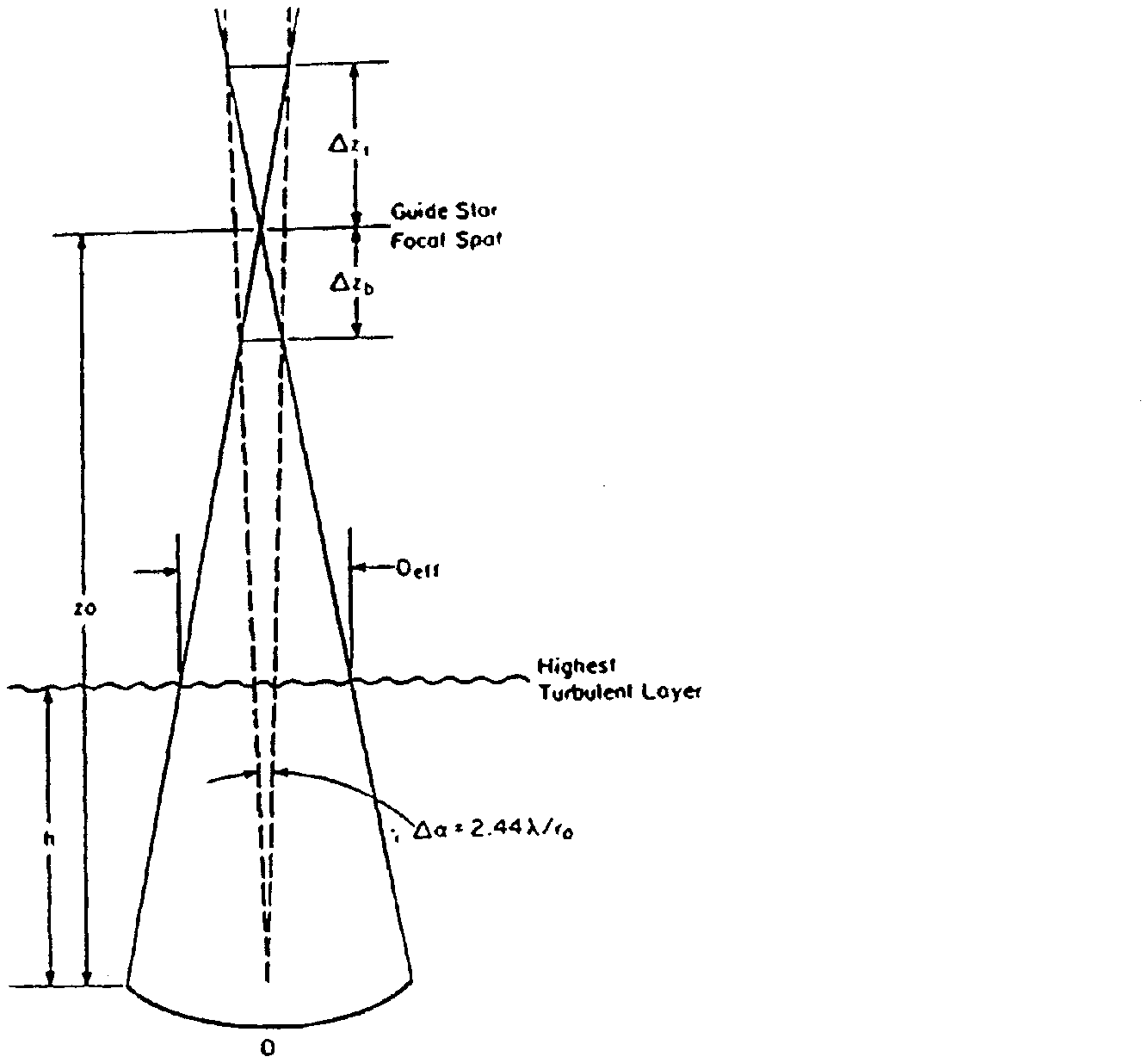}
\caption{Schematic representation of the laser projection geometry in a full aperture 
broadcast mode.  The telescope primary mirror is labeled D, but if it is partially illuminated D is 
replaced with $D_p$. The focus height of the laser beam is $z_o$ , and the distances from $z_o$ to the 
top and bottom of the Rayleigh scattered return region are, respectively,  $\Delta$$z_t$ and $\Delta$$z_b$.
\label{fig3}}
\end{figure*}  

\clearpage
\begin{figure*}
\epsscale{0.5}
\plotone{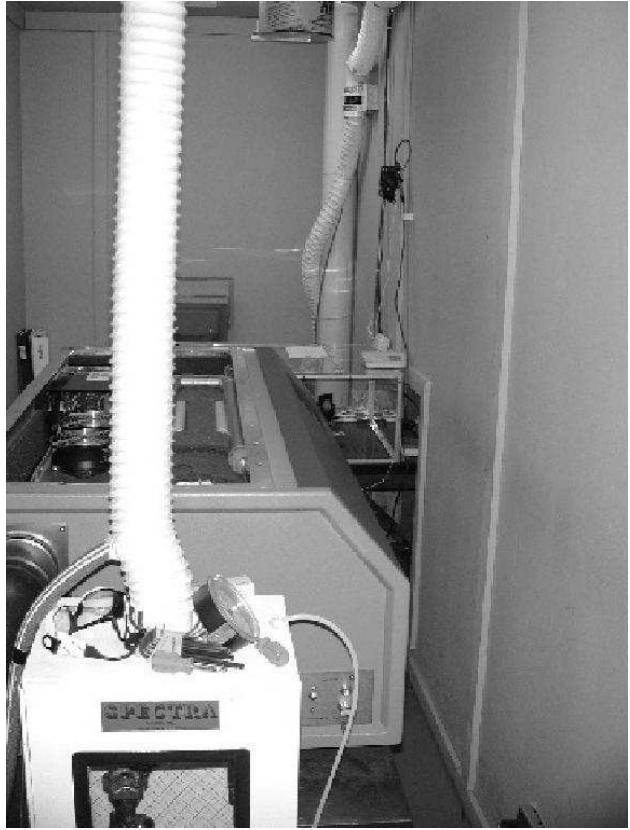}
\caption{Photograph of the Questek laser with the top cover removed.  In the foreground 
is a sealed cabinet holding the fluorine gas with a vent hose exiting the top of the cabinet.  
The chamber that holds the lasing gas mixture is just to the right of the vent hose and the 
laser capacitor banks are below the fans visible on the left of the vent hose.  The laser 
beam exits on the far side of the laser cabinet.
\label{fig4}}
\end{figure*}

\clearpage
\begin{figure*}
\epsscale{0.5}
\plotone{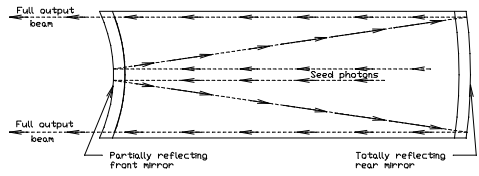}
\caption{Schematic drawing showing how photon amplification occurs in an unstable 
resonator cavity.
\label{fig5}}
\end{figure*}

\clearpage
\begin{figure*}
\epsscale{0.5}
\plotone{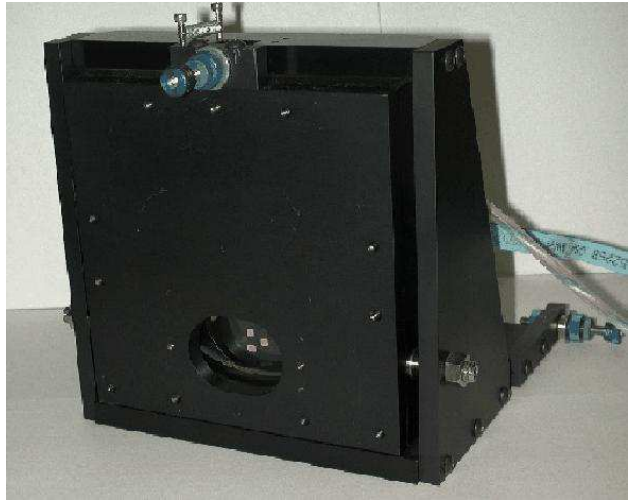}
\caption{UnISIS rotating disk assembly was removed from the optical system for this picture. A set of 3
reflective spots can be through the front face of the mounting.
\label{fig6}}
\end{figure*}

\clearpage
\begin{figure*}
\epsscale{0.5}
\plotone{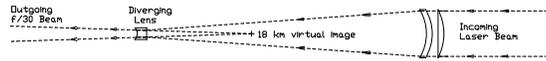}
\caption{Three simple lenses are used to convert the (nearly) collimated laser beam into f/30 for projection
with the 2.5m telescope optics. The ``air-spaced doublet'' is on the right and the ``diverging lens'' on the
left. These lenses represent surfaces 2 through 14 in Table 1. Note that the design has no concave surfaces
with radii on the right, a necessary condition to avoid back-reflected laser hot spots. The point labled
``18km virtual image'' is placed at the 18km conjugate point in the 2.5m telescope system to guarantee 
that the laser guide star focuses at the pre-selected altitude. 
\label{fig7}}
\end{figure*}

\clearpage
\begin{figure*}
\epsscale{0.5}
\plotone{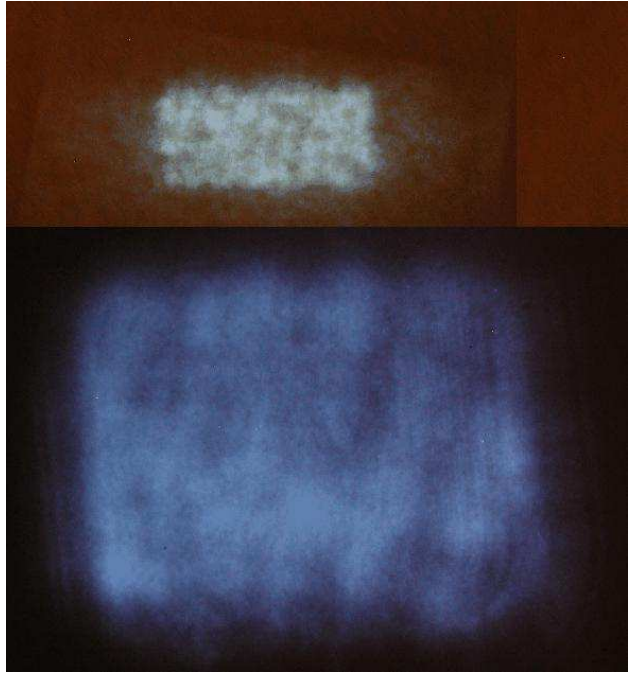}
\caption{  (Top)  The top panel shows the excimer laser beam before it encounters the razor 
beam stop.  At this point ($\sim$9 meters from the laser exit port) the beam is approximately 14 
mm x 28 mm.  (Bottom) The bottom panel shows the laser beam as it exits the Coude room on its 
way up the south polar axis.  This beam is $\sim$55 mm across the diagonal at this point.  
Interference fringes are visible in both images, but the peak intensity in the center of the 
fringes exceeds the uniform background only by $\sim$10\%.  The fringes are intrinsic to the 
laser and probably arise from interference phenomena within the front and or back laser 
mirrors.  The laser mirrors have a dielectric coating deposited on their external face (to 
prevent fluorine gas from attacking the dielectric material), so lasing photons pass through 
the glass at both ends of the laser chamber.
\label{fig8}}
\end{figure*}

\clearpage
\begin{figure*}
\epsscale{0.5}
\plotone{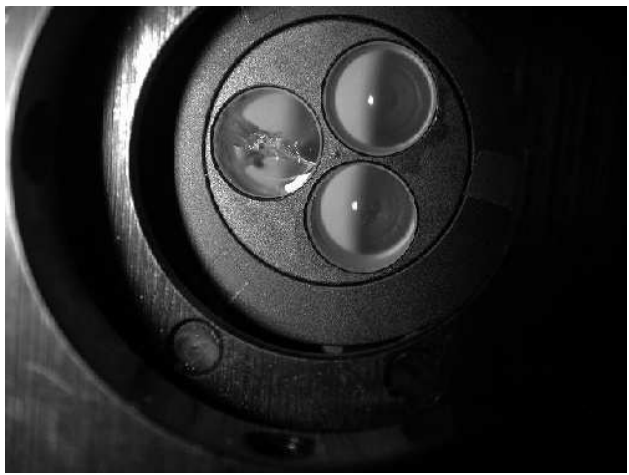}
\caption{Greatly enlarged image of the diverging lens holder.  While three lenses are 
held by a single mount, only one is used at a time.  Two pristine diverging lenses can be 
seen on the right, and the one damaged diverging lens is on the left.  The damage occurred 
when the energy density sent into the lens exceeded 6 J cm$^{-2}$ at the rear surface of 
the lens.  To set a scale for this image, each lens has a diameter of 7.4 mm.  
\label{fig9}}
\end{figure*}

\clearpage
\begin{figure*}
\epsscale{0.5}
\plotone{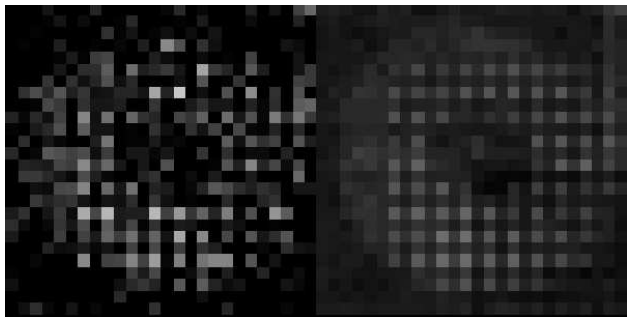}
\caption{UnISIS UV wavefront sensor images from the 18 km laser guide star.  (Left) the 
wavefront image from a single laser pulse.  (Right) the integrated wavefront signal from  45 
consecutive laser pulses.  As described in the text, the lenslet array was purposely offset to 
the lower left to show a portion of the bare pupil illuminated with the 351 nm laser light.
\label{fig10}}
\end{figure*}

\end{document}